\documentclass[10pt,conference]{IEEEtran}
\usepackage[utf8]{inputenc}
\IEEEoverridecommandlockouts
\usepackage{algorithmic}
\usepackage{algorithm}
\usepackage{bm}
\usepackage{cite}
\usepackage{amsmath,amssymb,amsfonts}
\usepackage{algorithmic}
\usepackage{graphicx}
\usepackage{textcomp}
\usepackage{xcolor}
\usepackage{makecell}
\usepackage{subfigure}
\usepackage{tcolorbox}
\usepackage{balance}

\usepackage{multirow}
\usepackage{diagbox}
\usepackage{graphicx}
\usepackage{booktabs}

\usepackage{xcolor}
\usepackage{hyperref}

\definecolor{citegreen}{RGB}{0,255,0}

\usepackage{etoolbox}
\makeatletter
\patchcmd{\@cite}
  {\hbox{[#1]}}
  {\fboxsep=1pt\colorbox{citegreen}{\hbox{[#1]}}}
  {}
  {}
\makeatother
\def\BibTeX{{\rm B\kern-.05em{\sc i\kern-.025em b}\kern-.08em
    T\kern-.1667em\lower.7ex\hbox{E}\kern-.125emX}}
\begin{document}

% \title{SAEL: Leveraging Large Language Models with Adaptive Mixture-of-Experts for Smart Contract Vulnerability Detection}
% \author{\IEEEauthorblockN{Anonymous Authors}}

\title{SAEL: Leveraging Large Language Models with Adaptive Mixture-of-Experts for Smart Contract Vulnerability Detection\\
% {\footnotesize \textsuperscript{}}
\thanks{$^{1}$Lei Yu and Shiqi Cheng contributed equally to this work.}
\thanks{$^{\ast}$Corresponding authors: Li Yang and Fengjun Zhang.}
}

\author{
  \IEEEauthorblockN{
      Lei Yu\IEEEauthorrefmark{2}\IEEEauthorrefmark{3}\textsuperscript{1},
      Shiqi Cheng\IEEEauthorrefmark{2}\textsuperscript{1},
      Zhirong Huang\IEEEauthorrefmark{2}\IEEEauthorrefmark{3},
      Jingyuan Zhang\IEEEauthorrefmark{2}\IEEEauthorrefmark{3},
      Chenjie Shen\IEEEauthorrefmark{2}\IEEEauthorrefmark{3}, \\
      Junyi Lu\IEEEauthorrefmark{2}\IEEEauthorrefmark{3}, 
      Li Yang\IEEEauthorrefmark{2}\textsuperscript{$\ast$},
      Fengjun Zhang\IEEEauthorrefmark{2}\IEEEauthorrefmark{4}\textsuperscript{$\ast$},
      Jiajia Ma\IEEEauthorrefmark{2}
    }
  \IEEEauthorblockA{
    \IEEEauthorrefmark{2}Institute of Software, Chinese Academy of Sciences, Beijing, China \\
    \IEEEauthorrefmark{3}University of Chinese Academy of Sciences, Beijing, China \\
    \IEEEauthorrefmark{4}State Key Laboratory of Computer Science, Institute of Software, Chinese Academy of Sciences, Beijing, China \\
    \{yulei2022, chengshiqi, huangzhirong2022, zhangjingyuan2023, lujunyi2022\}@iscas.ac.cn, 
    \\
    shenchenjie22@mails.ucas.ac.cn, \{yangli2017, fengjun, majiajia\}@iscas.ac.cn 
  }
}

\maketitle

\begin{abstract}

With the increasing security issues in blockchain, smart contract vulnerability detection has become a research focus. Existing vulnerability detection methods have their limitations: 1) Static analysis methods struggle with complex scenarios. 2) Methods based on specialized pre-trained models perform well on specific datasets but have limited generalization capabilities. In contrast, general-purpose Large Language Models (LLMs) demonstrate impressive ability in adapting to new vulnerability patterns. However, they often underperform on specific vulnerability types compared to methods based on specialized pre-trained models. We also observe that explanations generated by general-purpose LLMs can provide fine-grained code understanding information, contributing to improved detection performance. 

Inspired by these observations, we propose SAEL, a LLM-based framework for smart contract vulnerability detection. First, we design prompts targeting specific smart contract vulnerabilities to guide general-purpose LLMs in detecting vulnerabilities and providing explanations. The detection results generated by LLMs serve as prediction features. Then, we employ prompt-tuning on CodeT5 and T5 respectively to process contract code and explanations, enhancing model performance on specific tasks. To leverage the strengths of each component, we introduce Adaptive Mixture-of-Experts, a dynamic architecture for smart contract vulnerability detection. This mechanism dynamically adjusts feature weights through a Gating Network, which selects the most relevant features by applying TopK filtering and Softmax normalization, and a Multi-Head Self-Attention mechanism, which enhances cross-feature relationships by processing multiple attention heads in parallel. This design ensures that prediction results for LLMs, explanation features, and contract code features are effectively integrated through gradient optimization. The loss function focuses on the independent prediction performance of each feature and the overall performance of weighted predictions. Experimental results show that SAEL outperforms existing methods in detecting various vulnerabilities.
\end{abstract}

% To leverage the strengths of each component, we introduce Adaptive Mixture-of-Experts, a dynamic architecture for smart contract vulnerability detection. This mechanism dynamically adjusts feature weights through a Gating Network, which selects the most relevant features by applying TopK filtering and Softmax normalization, and a Multi-Head Self-Attention mechanism, which enhances cross-feature relationships by processing multiple attention heads in parallel. This design ensures that prediction results for LLMs, explanation features, and contract code features are effectively integrated through gradient optimization.


\section{Introduction}
Blockchain technology has grown rapidly and become popular across various sectors due to its decentralized nature \cite{swan2015blockchain}. As a significant innovation, blockchain allows for the creation of secure, decentralized, and distributed digital ledgers that record transactions \cite{hewa2021survey}. By using cryptographic methods, blockchain ensures that each transaction is secure and verified, making it a highly dependable technology \cite{wood2014ethereum, yu2023money}. In particular, graph analysis techniques, including detecting criminal communities \cite{yang2023dccgraph} and multi-scale anomaly detection \cite{zhang2025topology}, play a crucial role in identifying irregular patterns and enhancing the security of blockchain networks. Smart contracts are automated programs on the blockchain that let developers create rules for managing digital assets like cryptocurrency. They automatically execute when specific conditions are met. Once these programs are deployed on the blockchain, they are permanent \cite{zou2019smart}. However, the unchangeable and complex nature of smart contracts means that their security challenges are increasingly evident \cite{zou2019smart}. The infamous DAO attack \cite{dhillon2017dao,mehar2019understanding, yu2024smart, yu2025smart, yuan2025mos} illustrates the severe consequences of such vulnerabilities. This attack led to the unauthorized transfer of Ethereum worth 60 million dollars, causing significant disruptions in the blockchain community \cite{alharby2017blockchain, hegedHus2018towards}. This highlights the critical need to enhance the security of smart contracts to avoid such crushing outcomes in the future.

\begin{figure*}[!h]
\centerline{\includegraphics[width=0.8\textwidth,height=0.4\textwidth]{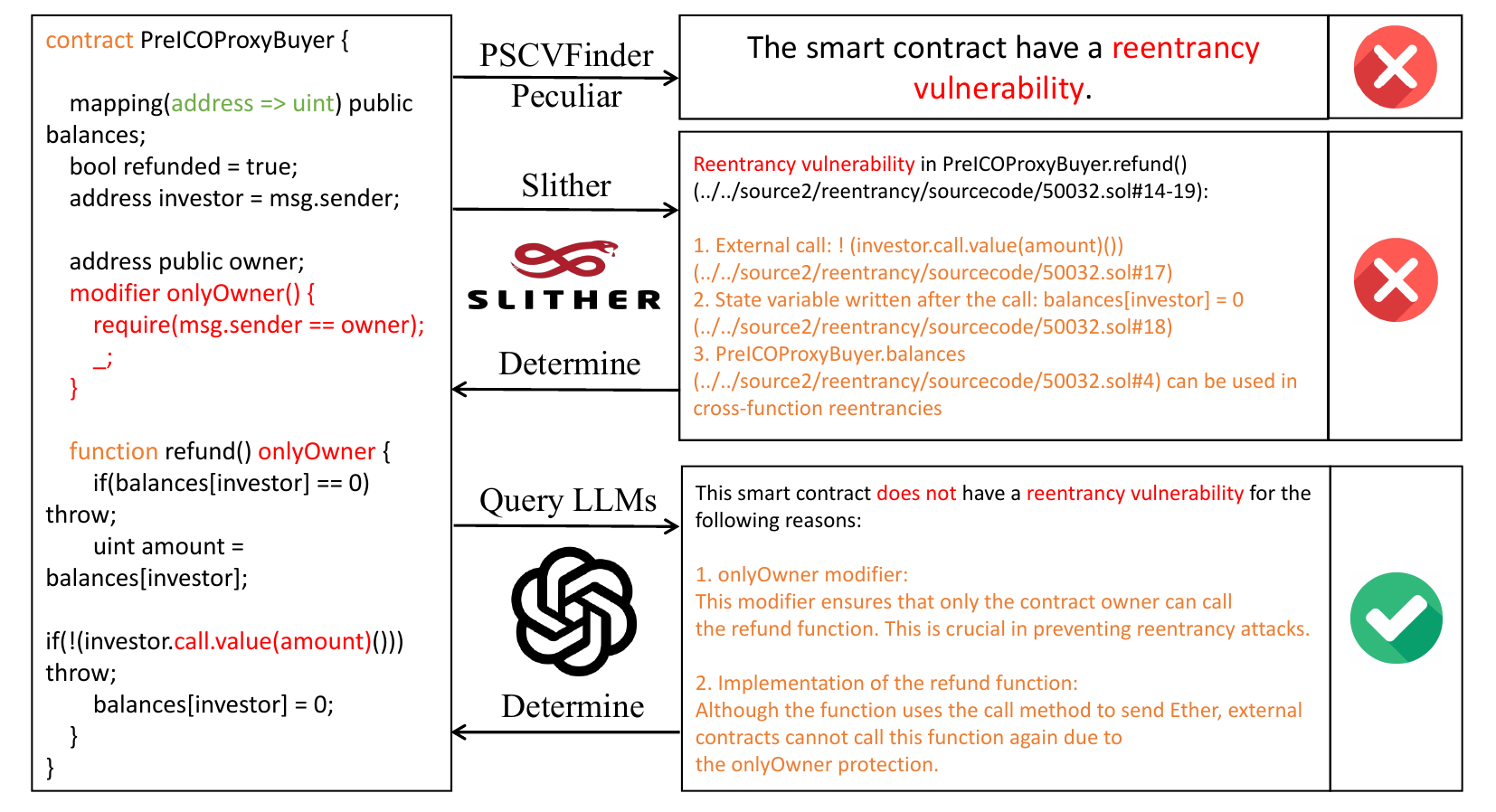}}
\caption{Motivation examples of different methods on smart contract vulnerability detection task.}
\label{motivation}
\end{figure*}

Researchers have developed various techniques to identify vulnerabilities in smart contracts. One popular approach is symbolic execution, which is implemented in tools such as Oyente \cite{luu2016making}, Mythril \cite{mueller2017mythril}, Osiris \cite{torres2018osiris}, and Manticore \cite{mossberg2019manticore}. Another commonly used technique is static analysis, which is employed by tools like Slither \cite{feist2019slither} and SmartCheck \cite{tikhomirov2018smartcheck}. However, these methods rely solely on fixed patterns and have poor generalizability. As shown in Fig. \ref{motivation}, Slither incorrectly identifies a reentrancy vulnerability by detecting external calls and state variable modifications, but fails to consider the onlyOwner modifier that prevents such attacks, resulting in a false positive. Clear \cite{chen2024improving} adopts a Contrastive Learning (CL) model to learn complex relationships between contracts. Zhuang et al. \cite{zhuang2020smart} and Luo et al. \cite{luo2024scvhunter} introduce a graph neural network-based approach that converts smart contract code into graph representations. However, the complexity of the graph structures employed in these techniques makes them difficult to reproduce and effectively represent programs. Peculiar \cite{wu2021peculiar} and PSCVFinder \cite{yu2023pscvfinder} achieve precise detection of smart contract vulnerabilities by fine-tuning pre-trained models. As shown in Fig. \ref{model_comparison}, their detection capabilities for reentrancy and timestamp dependency vulnerabilities are superior to directly using general-purpose LLMs. However, they may struggle to handle scenarios not well-represented in the training data, exhibiting poor generalizability. As illustrated in Fig. \ref{motivation}, specialized pre-trained models (PSCVFinder and Peculiar) fail to understand the role of the onlyOwner modifier in Fig. \ref{motivation}, resulting in false positives. In contrast, general-purpose LLM correctly understands the function of the onlyOwner modifier, recognizing that the refund function is protected by the onlyOwner modifier. Even if investor.call.value(amount)() triggers a callback function of a malicious contract, that malicious contract cannot call the refund function again because it is not the contract owner. Consequently, general-purposed LLM correctly concludes that the contract does not contain a reentrancy vulnerability. This comparison highlights the complementary strengths of different approaches: general-purpose LLMs demonstrate superior capability in adapting to new vulnerability patterns, while specialized pre-trained models provide high performance for well-defined vulnerability types.

We observe that explanations generated by general-purpose LLMs can serve as fine-grained code understanding information, improving the performance of smart contract vulnerability detection. As shown in Fig. \ref{timestamp}, for a specific contract, the general-purpose LLM (Qwen1.5-72B-Chat) identified a timestamp dependency vulnerability in the contract and provided explanations about the usage of block.timestamp (only used in a require statement), the structure and security checks of the function  (constrained by two require statements), and the usage of state variables (read-only). These details allow for the analysis that the timestamp is only used for simple comparison, does not directly affect the state, and state variables are not modified. These inferences based on detailed explanations suggest that the contract may not actually contain a timestamp dependency vulnerability, which is consistent with the label. This insight motivates us to incorporate the explanation generated by general-purpose LLMs into smart contract vulnerability detection models.

Based on these findings, we propose SAEL, an LLM-based smart contract vulnerability detection framework. First, we design prompts tailored to specific smart contract vulnerabilities to guide LLMs in analyzing smart contract code, detecting vulnerabilities, and providing fine-grained explanations. This addresses the poor generalizability of static analysis and pre-trained methods. The predictions generated by LLMs serve as predictive features. We conducted a comprehensive empirical study comparing the performance of different LLMs on reentrancy and timestamp dependency vulnerability datasets, as shown in Fig. \ref{model_comparison}. We found that Qwen1.5-72B-Chat exhibits performance closest to GPT-4-Turbo in smart contract vulnerability detection. To balance performance and overhead, we employ it as the LLM model in this study. Yu et al. \cite{yu2023pscvfinder} demonstrated that using prompt tuning outperforms fine-tuning on smart contract vulnerability detection tasks, while CodeT5 \cite{wang2021codet5} achieves better results compared to other models such as GraphCodeBERT \cite{guo2020graphcodebert}. Therefore, we adopt prompt-tuning instead of fine-tuning on CodeT5 \cite{wang2021codet5} and T5 \cite{raffel2020exploring} to handle contract code and explanations, respectively. Finally, to fully leverage the advantages of each component and effectively incorporate explanations into the detection model, we introduce Adaptive Mixture-of-Experts, a dynamic framework that integrates raw code features, LLM-generated explanations, and LLM predictions. Unlike static ensemble methods, Adaptive Mixture-of-Experts dynamically adjusts feature weights through a Gating Network, which selects the most relevant features by applying a TopK mechanism to filter the most significant feature dimensions, followed by Softmax normalization to produce a gating vector. In addition, a Multi-Head Self-Attention mechanism captures complex cross-dimensional relationships among the features by computing attention scores across multiple heads and combining them to enhance contextual understanding. These components ensure robust and adaptive detection by optimizing a learnable loss function through gradient descent, allowing the model to adaptively balance the contributions of different feature types. The loss function focuses on the independent predictive performance of each feature and the overall performance of the weighted predictions. By minimizing the loss function, we can obtain the optimal combination of feature weights. 

% To fully harness the strengths of each component, our approach enables effective integration of the explanation generated by LLMs into the smart contract vulnerability detection model, achieving precise and comprehensive vulnerability detection.

% We evaluated our SAEL framework on over 200,000 real-world smart contracts in the SmartBugs Wild dataset \cite{ferreira2020smartbugs} for reentrancy vulnerabilities and the ESC dataset \cite{liu2021smart} containing 40,932 Ethereum smart contracts and 307,396 functions for timestamp dependency vulnerabilities. For integer overflow/underflow and delegatecall vulnerabilities, we select two largest publicly available vulnerability dataset for
% smart contracts \cite{liu2023rethinking,qian2023cross} and mix them, which consists of more than 52K real-world smart contracts. The results demonstrate that SAEL outperforms state-of-the-art methods, with F1 scores 2.33\%, 3.16\%, 10.67\%, and 13.32\% higher for reentrancy, timestamp dependency, integer overflow/underflow, and delegatecall vulnerabilities, respectively. Moreover, SAEL exhibits strong zero-shot capabilities across the four vulnerability types.

We evaluated our SAEL framework on over 200,000 real-world smart contracts in the SmartBugs Wild dataset \cite{ferreira2020smartbugs} for reentrancy vulnerabilities and the ESC dataset \cite{liu2021smart}. For integer overflow/underflow and delegatecall vulnerabilities, we select two largest publicly available vulnerability dataset for smart contracts \cite{liu2023rethinking,qian2023cross} and mix them. The results demonstrate that SAEL outperforms state-of-the-art methods, with F1 scores 2.33\%, 3.16\%, 10.67\%, and 13.32\% higher for reentrancy, timestamp dependency, integer overflow/underflow, and delegatecall vulnerabilities, respectively. Moreover, SAEL exhibits strong zero-shot capabilities across the four vulnerability types.

The main contributions of this paper are as follows:
\begin{itemize}

\item We are among the first to incorporate LLM-generated explanations as features to enhance smart contract vulnerability detection, demonstrating their impact on improving detection performance.

\item To fully harness the strengths of each feature, we introduce Adaptive Mixture-of-Experts for smart contract vulnerability detection. It dynamically adjusts the weights assigned to the prediction results based on different features.

\item Our approach sets new state-of-the-art performance in detecting reentrancy, timestamp dependency, integer overflow/underflow, and delegatecall vulnerabilities in smart contracts.

\end{itemize}

% Due to institutional and confidentiality reasons, the source code associated with this work are not publicly available at this time.
All source code and data in this study are publicly available at \cite{yu2025sael}.

\section{Background and Motivation}

\subsection{Problem Statement}

We propose an automated approach to detect vulnerabilities in individual smart contract functions. Our method assigns a label $\hat{y}$ to each function f, where $\hat{y}=1$ indicates a vulnerability and $\hat{y}=0$ denotes security. We focus on four key vulnerability types.

\textbf{Reentrancy vulnerability} occurs when a contract calls an external contract or sends Ether before completing all necessary internal state changes. An attacker can exploit this vulnerability by repeatedly calling the vulnerable function before the original call is completed, potentially leading to unexpected behavior such as multiple withdrawals of funds.

\textbf{Timestamp dependence vulnerability} occurs when smart contracts rely on block timestamps for critical operations. Miners can manipulate these timestamps, potentially compromising contract integrity and leading to financial losses. This vulnerability often affects contracts using timestamps for random number generation or key decision-making processes.

\textbf{Integer Overflow/Underflow} occurs when the result of an arithmetic operation exceeds the storage range of the variable. In an overflow, the value "wraps around" to the minimum value for that type, while in an underflow, it "wraps around" to the maximum value. This can lead to unexpected contract behavior such as incorrect balances or out-of-control loops.

\textbf{Delegatecall} is a low-level function call that allows a contract to dynamically load code from another contract. While this provides powerful upgradeability, it can lead to severe security vulnerabilities if used improperly. The main risk is that the called contract executes in the context of the calling contract and can thus modify the calling contract's storage.

\textbf{We primarily focus on these four vulnerabilities for the following reasons:} (i) Empirical evidence shows that approximately 70\% of financial losses in Ethereum smart contract attacks are attributed to these vulnerabilities \cite{chen2020survey}. (ii) Exisiting works \cite{chen2020survey,gao2019easyflow,praitheeshan2019security} demonstrates that these vulnerabilities occur with higher frequency in Ethereum smart contracts compared to others.

\begin{figure}[htbp]
\centerline{\includegraphics[width=0.48\textwidth]{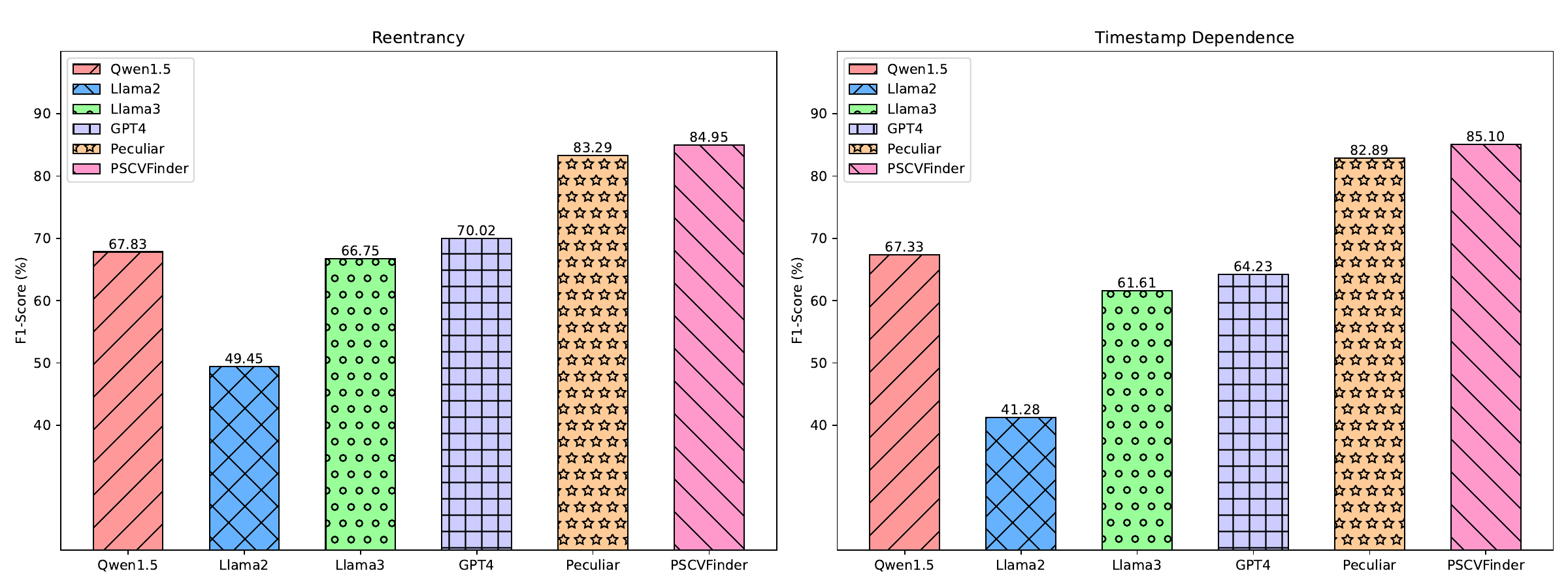}}
\caption{Performance comparison of specialized pre-trained models and general-purpose LLMs.}
\label{model_comparison}
\end{figure}

\begin{figure}[htbp]
\centerline{\includegraphics[width=0.5\textwidth]{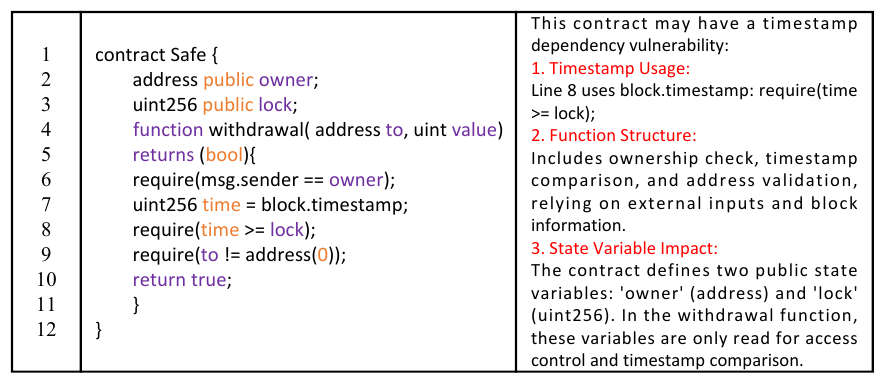}}
\caption{The case related to timestamp where LLMs made incorrect predictions.}
\label{timestamp}
\end{figure}

\begin{figure*}[!h]
\centerline{\includegraphics[width=0.9\textwidth,height=0.45\textwidth]{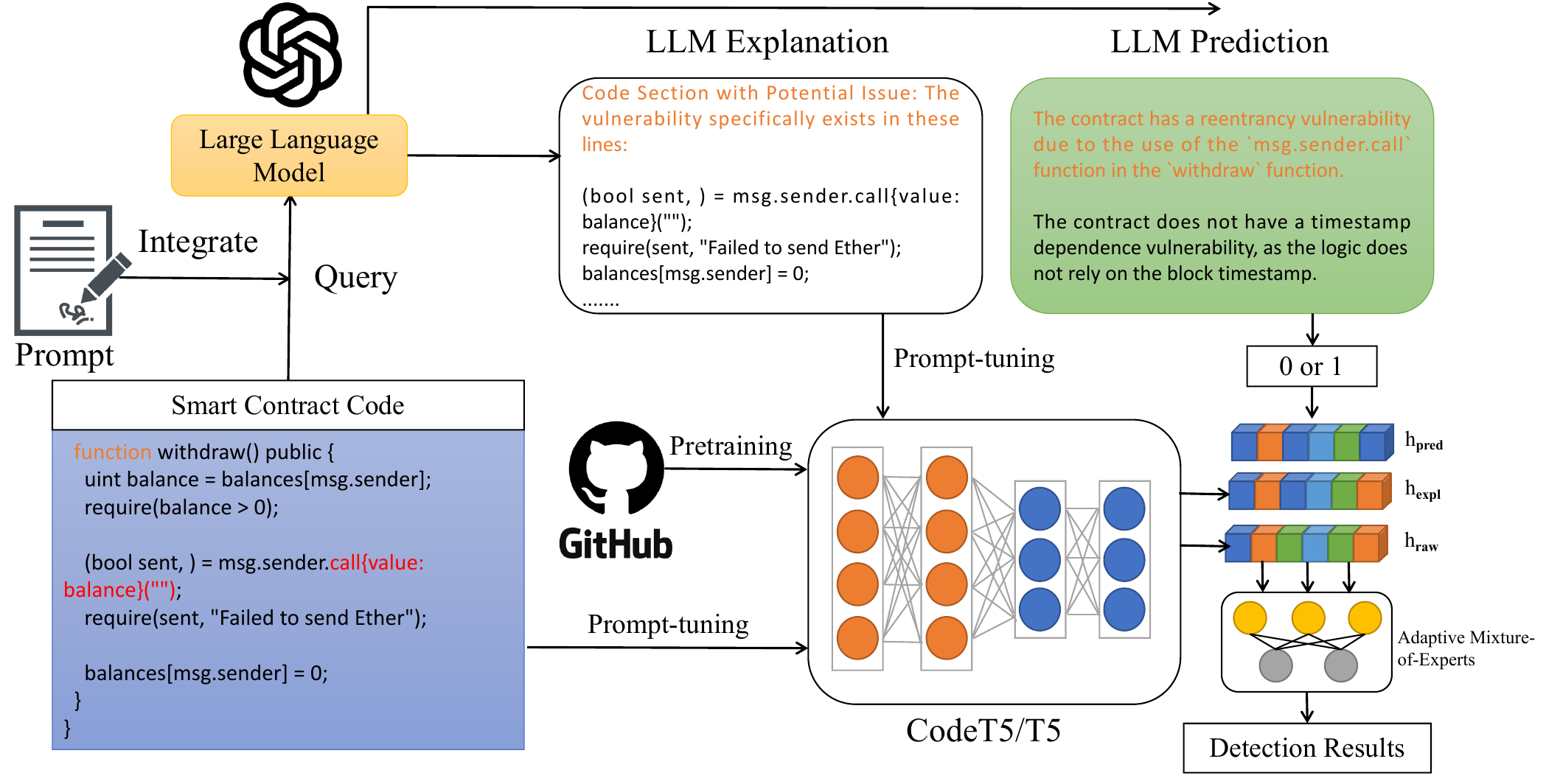}}
\caption{The overall architecture of the proposed model SAEL.}
\label{overview}
\end{figure*}

\subsection{Motivating Examples}

In this section, we use two real-world smart contract examples to illustrate the complementary strengths of different approaches in detecting smart contract vulnerabilities: specialized pre-trained models fine-tuned on specific vulnerability datasets, general-purpose Large Language Models (LLMs), and the explanations generated by LLMs.

\textbf{Example 1: Comparative Analysis Reveals Distinct Strengths of Different Approaches.} Static analysis methods (e.g., Slither) rely on predefined vulnerability patterns and may struggle with complex scenarios. As shown in Fig. \ref{motivation}, Slither incorrectly identifies a reentrancy vulnerability by detecting external calls and state variable modifications, without considering the onlyOwner modifier that prevents such attacks. Specialized pre-trained models (e.g., PSCVFinder and Peculiar), fine-tuned on specific vulnerability datasets, demonstrate high performance in detecting known vulnerability types, often outperforming general-purpose LLMs on these specific tasks, as illustrated in Fig. \ref{model_comparison}. However, they may struggle with scenarios not well-represented in their training data, showing poor generalization and failing to understand the onlyOwner modifier in Fig. \ref{motivation}. In contrast, general-purpose LLMs correctly comprehend the role of the onlyOwner modifier, leading to the accurate conclusion that the smart contract does not contain a reentrancy vulnerability. This comparison highlights the complementary strengths of different approaches: general-purpose LLMs offer superior capability in adapting to new vulnerability patterns, while specialized pre-trained models provide high performance for well-defined vulnerability types.

\textbf{Example 2: Explanations Generated by LLMs Can Infer the Correct Answer.} Our case studies reveal that explanations generated by general-purpose Large Language Models (LLMs) can provide valuable insights for improving smart contract vulnerability detection. For example, consider the contract in Fig. \ref{timestamp}, the LLM incorrectly states that the contract may have a timestamp dependency vulnerability. Despite predicting a timestamp dependency vulnerability in the smart contract, the LLM provided relatively comprehensive code analysis as shown in Fig. \ref{timestamp}: 1) It correctly identified the location where the timestamp is used. 2) It accurately described the structure and security checks of the function. 3) It correctly identified the state variables and how they are used in the function (only read, not modified). Based on the code analysis, we can infer that: 1) The timestamp is only used for simple comparison and does not directly affect the state. 2) The function includes multiple security checks. 3) State variables are not modified. These inferences based on detailed explanations suggest that the contract may not actually contain a timestamp dependency vulnerability, which is consistent with the ground truth. This example demonstrates how LLM-generated explanations can serve as a form of code analysis, potentially refining the initial prediction of the model.

Based on these examples, we observe that while specialized pre-trained models may outperform general-purpose LLMs in detecting specific, well-defined vulnerability types, general-purpose LLMs exhibit superior flexibility in adapting to new or complex vulnerability patterns. Moreover, the explanations generated by LLMs provide an additional layer of analysis that can enhance detection performance, even when the initial prediction is incorrect. Combining these approaches with LLM-generated explanations can complement each other and enhance detection performance.

\section{Approach}

The overall workflow of SAEL is shown in Fig. \ref{overview}. The SAEL framework consists of three key modules: the design of prompt templates, T5-based Prompt-tuning, and Adaptive Mixture-of-Experts (MoE). Each module is carefully designed to enhance the performance of smart contract vulnerability detection.

\subsection{The Design of Prompt Templates}

Smart Contract Vulnerability Detection methods \cite{hu2023large,chen2023chatgpt,david2023you} based on LLMs typically employ models like GPT-4-Turbo, but the cost of calling their APIs is high. To reduce costs while maintaining detection performance, we choose the open-source Qwen1.5-72B-Chat as the LLM model for our work. We evaluated the performance of Qwen1.5-72B-Chat and GPT-4-Turbo on smart contract vulnerability detection tasks and found that Qwen1.5-72B-Chat achieves comparable detection effectiveness with significantly lower computational cost, making it an ideal choice for our framework.

\begin{figure}[htbp]
\centerline{\includegraphics[width=0.48\textwidth]{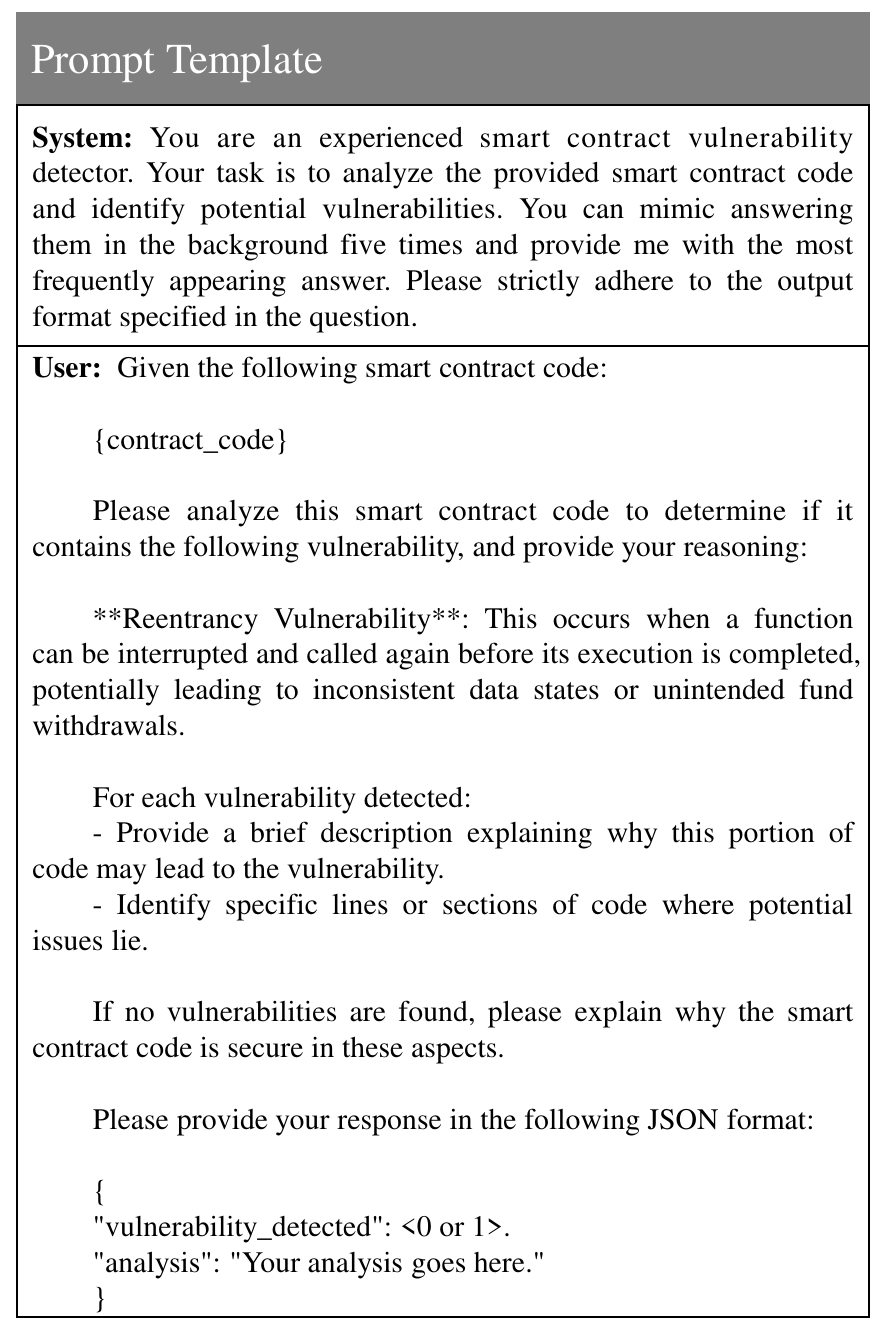}}
\caption{The Prompt Design for Reentrancy Vulnerability.}
\label{prompt}
\end{figure}

For four types of vulnerabilities (e.g., reentrancy, timestamp dependence), we carefully designed prompt templates. Unlike existing works \cite{hu2023large,chen2023chatgpt,david2023you}, which simply provide general vulnerability descriptions or assign the identity of a "smart contract security expert" to LLMs, our prompt templates are tailored for each specific vulnerability type. For instance, as shown in Fig. \ref{prompt}, the prompt template for detecting reentrancy vulnerabilities includes:
1. A detailed definition of reentrancy vulnerabilities and necessary background knowledge.
2. A description of typical characteristics associated with reentrancy vulnerabilities (e.g., external calls in loops).
3. Instructions for analyzing the given code, identifying vulnerabilities, explaining their causes, and locating problematic code sections.

To ensure structured and logical reasoning, the prompt adopts a Chain-of-Thought reasoning process, guiding the LLM step-by-step to: 
(1) understand the vulnerability's definition and characteristics, 
(2) analyze the code structure, 
(3) identify potential vulnerabilities, and 
(4) explain the causes or provide evidence of security. This process helps the LLM focus on key aspects of the code and ensures reliable detection results.

\textbf{Mimic-in-the-Background Approach:}  
To further enhance consistency, we employ the "mimic-in-the-background" approach \cite{sun2024gptscan}. Specifically, the LLM generates five responses for the same prompt in the background. The system then selects the most frequently appearing answer, ensuring the final response is both representative and reliable.

\subsection{T5-based Prompt-tuning}

To extract semantic information from smart contract code, we utilize two pre-trained models: the code language model CodeT5 \cite{wang2021codet5} and the text language model T5 \cite{raffel2020exploring}. CodeT5 processes raw smart contract code, while T5 processes natural language explanations generated by the LLM. Instead of fine-tuning, we employ prompt-tuning, which has been shown to outperform fine-tuning on smart contract vulnerability detection tasks \cite{yu2023pscvfinder}.

Prompt-tuning learns a continuous task-specific prompt prepended to the input, guiding the model to focus on relevant information. A cloze-style template $f_{\text{prompt}}(x)$ with an input slot $[X]$ and an answer slot $[Z]$ is designed as follows:
\begin{equation}
f_{\text{prompt}}(x) = \text{"[X] The code is [Z]"}
\end{equation}

A verbalizer $V$ maps label words to the predicted class:
\begin{equation}
V = \begin{cases} 
\text{Vulnerable:} & [\text{defective, bad}] \\ 
\text{Secure:} & [\text{clean, perfect}]
\end{cases}
\end{equation}

The outputs of CodeT5 and T5 are denoted as $h_{\text{raw}}$ and $h_{\text{expl}}$, respectively:
\begin{equation}
h_{\text{raw}} = \text{CodeT5}_{\text{raw}}(x^{\text{raw}}) \in \mathbb{R}^{N \times d}
\end{equation}
\begin{equation}
h_{\text{expl}} = \text{T5}_{\text{expl}}(x^{\text{expl}}) \in \mathbb{R}^{N \times d}
\end{equation}

Additionally, the LLM prediction outputs are encoded into one-hot vectors and transformed into uniform feature embeddings $h_{\text{pred}}$. These three feature representations ($h_{\text{raw}}$, $h_{\text{expl}}$, $h_{\text{pred}}$) serve as inputs to the next module.

\textbf{Feature Integration:}  
The outputs of CodeT5 ($h_{\text{raw}}$) and T5 ($h_{\text{expl}}$) are concatenated with the LLM prediction embeddings ($h_{\text{pred}}$) to form the combined feature vector $x = [h_{\text{raw}}, h_{\text{expl}}, h_{\text{pred}}]$. This vector serves as the input to the Adaptive Mixture-of-Experts module for dynamic expert selection.

\subsection{Adaptive Mixture-of-Experts}

To maximize the strengths of different feature types and improve the adaptability of the detection process, we introduce an Adaptive Mixture-of-Experts (MoE) architecture. This architecture dynamically selects and combines expert outputs using a gating network and multi-head self-attention mechanisms, ensuring that the most relevant features are utilized for smart contract vulnerability detection.

% \textbf{Gating Network Design.}  
% The Gating Network plays a critical role in dynamically assigning weights to each expert based on the combined input features. Specifically, the input vector $x$ is constructed by concatenating raw code embeddings ($h_{\text{raw}}$), explanation embeddings ($h_{\text{expl}}$), and LLM prediction embeddings ($h_{\text{pred}}$). The Gating Network processes the input vector with a multi-layer transformation pipeline, including:
% \begin{enumerate}
%     \item \textbf{Feature Extraction}: A linear layer reduces the dimensionality of the input vector $x$, ensuring computational efficiency.
%     \item \textbf{Key Feature Selection}: A TopK function is applied to retain only the top $k$ most significant features in the vector, improving focus on the most relevant information.
%     \item \textbf{Normalization}: The Softmax function normalizes the retained features, generating the gating vector $G(x)$, which dynamically determines the weight of each expert.
% \end{enumerate}

% The final formula for the gating vector $G(x)$ is:
% \begin{equation}
% G(x) = \text{Softmax}(\text{TopK}(H(x), k))
% \end{equation}
% where $H(x)$ represents the feature vector transformed by the Gating Network. The TopK function ensures that only the most important features contribute to the final gating vector.

\textbf{Gating Network Design.}  
The Gating Network plays a critical role in dynamically assigning weights to each expert based on the combined input features. Specifically, the input vector $x$ is constructed by concatenating raw code embeddings ($h_{\text{raw}}$), explanation embeddings ($h_{\text{expl}}$), and LLM prediction embeddings ($h_{\text{pred}}$). The Gating Network processes the input vector with a multi-layer transformation pipeline, including:
\begin{enumerate}
    \item \textbf{Feature Extraction}: A linear layer reduces the dimensionality of the input vector $x$, ensuring computational efficiency.
    \item \textbf{Key Feature Selection}: A TopK function with $k=3$ is applied to retain only the top 3 most significant features in the vector, improving focus on the most relevant information.
    \item \textbf{Normalization}: The Softmax function normalizes the retained features, generating the gating vector $G(x)$, which dynamically determines the weight of each expert.
\end{enumerate}

The final formula for the gating vector $G(x)$ is:
\begin{equation}
G(x) = \text{Softmax}(\text{TopK}(H(x), k=3))
\end{equation}
where $H(x)$ represents the feature vector transformed by the Gating Network. The TopK function with $k=3$ ensures that only the most important features contribute to the final gating vector. This specific value of $k$ was selected to align with our feature space, which comprises three primary types (raw code, explanations, and predictions), setting $k=3$ ensures each feature type has the opportunity to be considered in the vulnerability detection process.

\textbf{Multi-Head Self-Attention Mechanism.}  
To further enhance feature representation, we leverage a Multi-Head Self-Attention mechanism. This mechanism enables parallel processing of multiple attention heads, each focusing on a different subspace of the input vector. The steps are as follows:
\begin{itemize}
    \item Transform the input vector into query, key, and value matrices.
    \item Compute attention weights by applying the scaled dot product between queries and keys, followed by a Softmax normalization.
    \item Use the attention weights to compute a weighted sum of the value vectors.
    \item Concatenate outputs from all attention heads and pass them through a final linear transformation to restore the original dimension.
\end{itemize}

This mechanism ensures comprehensive exploration of input features while preserving cross-dimensional contextual relationships.

\textbf{Expert Models and Weighted Accuracy Matrix.}  
The MoE model includes three specialized expert models:
\begin{enumerate}
    \item \textbf{Raw Code Expert}: Focuses on syntactic and structural patterns in $h_{\text{raw}}$, such as loop structures and external function calls.
    \item \textbf{Explanation Expert}: Processes $h_{\text{expl}}$ to extract contextual insights from LLM-generated explanations.
    \item \textbf{Prediction Expert}: Processes $h_{\text{pred}}$ to capture high-level semantic patterns based on LLM predictions.
\end{enumerate}

Each expert outputs a vector $O_i$ representing its prediction confidence. The outputs from all experts are combined into an accuracy matrix $M$, where each row corresponds to an expert, and each column represents a vulnerability type:
\begin{equation}
M = [O_1, O_2, O_3]
\end{equation}
The gating vector $G(x)$ is then used to weight the accuracy matrix, producing a weighted accuracy matrix:
\begin{equation}
M_{\text{weighted}} = G(x) \cdot M
\end{equation}

\textbf{Final Prediction}  
For each vulnerability type, the expert tool with the highest weighted accuracy is selected to make the final prediction:

\begin{equation}
O_{\text{final}} = \sum_{i=1}^{3} G_i(x) \cdot O_i
\end{equation}

% This approach ensures that the most suitable expert contributes the most to the final prediction, improving both robustness and accuracy in vulnerability detection.

\textbf{Loss Function Optimization.}  
Our loss function comprises three components for optimizing the model, where $O_i$ means the output predicted with each individual feature and $O_{\text{final}}$ means the output of weighted features.

Feature Weight Adjustment Loss $L_{\text{feature}}$ focuses on the independent performance of each feature prediction, assessing their contribution through weighted cross-entropy loss:
\begin{equation}
L_{\text{feature}} = \sum_{i\in\{\text{raw}, \text{expl}, \text{pred}\}} w'_i \cdot L_{\text{cross entropy}}(O_i, Y)
\end{equation}

Overall Cross-Entropy Loss $L_{\text{pred}}$ measures the overall prediction performance for the weighted average prediction of all features:
\begin{equation}
O_{\text{final}} = \sum_{i\in\{\text{raw}, \text{expl}, \text{pred}\}} w'_i \cdot O_i
\end{equation}
\begin{equation}
L_{\text{pred}} = L_{\text{cross entropy}}(O_{\text{final}}, Y)
\end{equation}

Weight Regularization Loss $L_{\text{reg}}$ is designed to balance the loss functions to prevent rapid weight updates:
\begin{equation}
L_{\text{reg}} = \gamma(\|w'_{\text{raw}} - w_{\text{raw}}\|^2 + \|w'_{\text{expl}} - w_{\text{expl}}\|^2 + \|w'_{\text{pred}} - w_{\text{pred}}\|^2)
\end{equation}

The final loss function is a weighted combination of these components, where $\alpha$ denotes the balance coefficient that minimizes the average final loss:
\begin{equation}
L_{\text{total}} = \alpha \cdot (L_{\text{feature}} + L_{\text{reg}}) + (1 - \alpha) \cdot L_{\text{pred}}
\end{equation}

Weight Update: We use gradient descent based on the gradient of the total loss function to adaptively adjust weights during training:
\begin{equation}
w'_i := w'_i - \eta \cdot \frac{\partial L_{\text{total}}}{\partial w'_i}
\end{equation}

This optimization process ensures that our Adaptive Mixture-of-Experts dynamically adjusts the importance of each feature type according to different smart contracts, fully leveraging the strengths of each component to achieve optimal vulnerability detection performance.

\section{Experiments}
\subsection{Research Questions}
To evaluate our proposed SAEL approach, we conduct experiments to answer the following research questions:

$\bullet$ $\textbf{RQ1}$: How effective does our proposed model SAEL perform compared to state-of-the-art methods?

$\bullet$ $\textbf{RQ2}$: What is the contribution of various key components and different features in the proposed SAEL framework to its overall performance?

$\bullet$ $\textbf{RQ3}$: How do the parameters of SAEL affect the performance of the model?

$\bullet$ $\textbf{RQ4}$: Can SAEL identify smart contract vulnerabilities in a zero-shot manner?

% \bullet $\textbf{RQ1}: How effective does our proposed model SAEL perform compared to state-of-the-art methods?$

% \bullet $\textbf{RQ2}: What is the contribution of various key components and different features in the proposed SAEL framework to its overall performance?$

% \bullet $\textbf{RQ3}: How do the parameters of SAEL affect the performance of the model?$

% \bullet $\textbf{RQ4}: Can SAEL identify smart contract vulnerabilities in a zero-shot manner?$

\subsection{Dataset}

For the evaluation of our approach in detecting reentrancy vulnerabilities, we employ the recently introduced SmartBugs Wild Dataset \cite{ferreira2020smartbugs} as our benchmark. This comprehensive dataset comprises 47,398 distinct Solidity language files, encompassing a total of approximately 203,716 contracts with identified vulnerabilities.

To assess the effectiveness of our method in identifying timestamp dependency vulnerabilities, we make use of the ESC (Ethereum Smart Contracts) Dataset \cite{liu2021smart}. This dataset is composed of 40,932 Ethereum smart contracts and concentrates on two specific types of vulnerabilities: reentrancy and time dependence. The dataset includes a total of 307,396 functions, out of which nearly 4,833 functions contain the block.timestamp match pattern, which serves as a potential indicator of time dependence vulnerabilities. In our experiments, we specifically focus on the functions that exhibit the block.timestamp match pattern as our dataset.

For our study on integer overflow/underflow and delegatecall vulnerabilities, we have integrated two of the most comprehensive publicly available vulnerability datasets for smart contracts \cite{liu2023rethinking,qian2023cross}. 

% This consolidated dataset encompasses over 52,000 real-world smart contracts.
\subsection{Baselines}
In our evaluation, we first select a set of baselines specifically designed for Smart Contract Vulnerability Detection. They can be broadly classified into three categories: rule-based techniques, pre-trained models-based techniques and LLM-based techniques.

Baseline methods, categorized as rule-based techniques, employ predefined heuristics to detect vulnerabilities in smart contracts. This category includes tools such as Manticore \cite{mossberg2019manticore}, Mythril \cite{mueller2017mythril}, Osiris \cite{torres2018osiris}, Oyente \cite{luu2016making}, Slither \cite{feist2019slither}, Securify \cite{tsankov2018securify}, and Smartcheck \cite{tikhomirov2018smartcheck}.

Pre-trained models-based techniques, rely on pre-trained models like CodeT5 \cite{wang2021codet5}, CodeBERT \cite{feng2020codebert}, GraphCodeBERT \cite{guo2020graphcodebert} and fine-tuning techniques to identify smart contract vulnerabilities, including Peculiar \cite{wu2021peculiar}, PSCVFinder \cite{yu2023pscvfinder} and ReVulDL \cite{zhang2022reentrancy}.

LLM-based techniques, which rely on LLMs to identify vulnerabilties in smart contract, including GPTScan \cite{sun2024gptscan} and iAudit \cite{ma2024combining}.

% LLM-based techniques, rely on large language models like GPT-3.5 \cite{openai_gpt3.5_2023}, GPT-4-Turbo \cite{achiam2023gpt,openai_gpt4_2023} and other open-sourced LLMs to identify smart contract vulnerabilities, including Llama-2-70B-Chat \cite{touvron2023llama}, Llama-3-70B-Instruct, Qwen1.5-72B-Chat \cite{bai2023qwen}, GPT-3.5-Turbo \cite{openai_gpt3.5_2023} and GPT-4-Turbo \cite{achiam2023gpt,openai_gpt4_2023}.

\begin{table}
    \caption[]{Training Hyperparameters}
    \label{hyperparameter}
    \centering
    \begin{tabular}{c|c|c}
    \hline
        ~ & Hyperparameter & Value \\ \hline
            &   Max input length & 2048\\ 
            \textbf{Large Language Models} & Max output length & 512  \\ 
        \textbf{Inference} & Top-p & 1   \\ 
         & Temperature & 0   \\ 
         & Repetition penalty & 1.2  \\ 

         \hline
         & Learning rate & 5e-5  \\ 
       \textbf{CodeT5/T5} & Max input length & 512  \\ 
        \textbf{Training} & Max output length & 32   \\ 
         & Beam size & 10  \\ 
         & Batch size & 32 \\ 
\hline
    \end{tabular}
\end{table}

\subsection{Metrics}
To evaluate the performance of our proposed model and other baseline approaches in identifying smart, we employed widely accepted evaluation criteria, namely Precision, Recall, and F1-score. Precision measures the proportion of correctly identified vulnerabilities among all the predicted positive cases. Recall, on the other hand, represents the fraction of correctly detected vulnerabilities out of all the actual vulnerabilities present in the dataset. Lastly, the F1-score provides a balanced measure by calculating the harmonic mean between Precision and Recall, giving equal weight to both metrics.

\textbf{Why we choose these metrics.} In real-world scenarios, secure smart contracts significantly outnumber those with vulnerabilities, resulting in imbalanced datasets. Under such circumstances, accuracy measures may yield misleading results. In contrast, the three metrics can more accurately reflect model performance on imbalanced data.

\begin{table*}[!ht]
    \caption{The performance of our method compared with 12 baselines in terms of Precision, Recall and F1-score.}
    \label{comparison}
    \centering
    % \begin{tabular}{@{}c|cccc|cccc|cccc|cccc@{}}
    \begin{tabular}{@{}c@{\hspace{1pt}}|cccc@{\hspace{2pt}}|cccc@{\hspace{2pt}}|cccc@{\hspace{2pt}}|cccc@{}}
        \toprule
        \raisebox{-1\height}{\centering Methods} & \multicolumn{4}{c|}{Reentrancy} & \multicolumn{4}{c|}{Timestamp Dependency} & \multicolumn{4}{c|}{Overflow/Underflow} & \multicolumn{4}{c}{Delegatecall}\\ 
        & P(\%) & R(\%) & F1(\%)& Rank  & P(\%) & R(\%) & F1(\%) & Rank & P(\%) & R(\%) & F1(\%)& Rank& P(\%) & R(\%) & F1(\%) & Rank\\ 
        \midrule
        Manticore & 50.00 & 50.36 & 50.18  & 13 & --& --& --& -- & -- & --  & -- & -- & -- & -- & -- & --\\ 
        Mythril & 50.35 & 51.80 & 51.06 & 12& 50.00 & 41.79 & 45.53  & 7 & 25.30 & 46.67 & 32.81& 10 & 42.99 & 74.19 & 54.44 & 5 \\ 
        Osiris & 59.06 & 53.96 & 56.39 & 10 & 52.41 & 36.71 & 43.18 & 8 & 45.33 & 75.56 & 56.57 & 5 & --& -- & -- & --\\ 
        Oyente & 65.79 & 53.96 & 59.29 & 8 & 45.17 & 38.41 & 41.51 & 9 & 60.87 & 46.67 & 52.83 & 6 & 40.43 & 30.65 & 34.86 & 9 \\ 
        Slither & 52.00 & 65.47 & 57.96 & 9 & 67.26 & 72.46 & 69.77 & 4 & 32.28 & 45.56 & 37.79 & 8 & 39.04 & 91.94 & 54.81 & 4\\ 
        Securify & 52.78 & 54.68 & 53.71  & 11 & -- & --& -- & -- & -- & -- & -- & -- & -- & -- & -- & --\\ 
        Smartcheck & 77.87 & 68.35 & 72.80 & 6 & 39.24 & 37.44 & 38.32 & 10 & 31.25 & 38.89 & 34.65 & 9 & 32.93 & 43.55 & 37.50 & 8\\ 
        \midrule
        Peculiar & 89.13 & 88.49 & 88.81  & 4 & -- & -- & -- & -- & 74.73 & 75.56 & 75.14 & 2 & 66.67 & 61.29 & 63.87 & 3\\ 
        ReVulDL & 91.49 & 92.81 & 92.14 & 2 & 88.09 & 85.75 & 86.90 & 3 & -- & -- & -- & -- & -- & -- & -- & --\\ 
        PSCVFinder & 92.65 & 90.65 & 91.64 & 3 & 90.64 & 88.89 & 89.76 & 2 & 65.00 & 72.22 & 68.42 & 3 & 69.23 & 72.58 & 70.87 & 2\\ 
        \midrule
        GPTScan & 62.22 & 80.58 & 70.22 & 7 & 57.41 & 74.88 & 64.99 & 6 & 39.11 & 77.78 & 52.04 & 7 & 31.43 & 88.71 & 46.41 & 6\\ 
        iAudit & 65.79 & 89.93 & 75.99 & 5 & 57.07 & 84.78 & 68.22 & 5 & 42.86 & 83.33 & 56.60 & 4 & 28.80 & 85.48 & 43.09 & 7\\ 
        \midrule
        \textbf{SAEL} & \textbf{93.62} & \textbf{94.96} & \textbf{94.29} & \textbf{1} & \textbf{90.16} & \textbf{95.17} & \textbf{92.60} & \textbf{1} & \textbf{79.00} & \textbf{87.78} & \textbf{83.16} & \textbf{1} & \textbf{78.46}  & \textbf{82.26} & \textbf{80.31} & \textbf{1}\\
        \bottomrule
    \end{tabular}
\end{table*}

\subsection{Implementation Details}

We leveraged CodeT5 \cite{wang2021codet5} and T5 \cite{raffel2020exploring} which followed the initialization of its pre-training work. As shown in Table \ref{hyperparameter}, for the training of CodeT5 and T5, we set the max input length to 512, the max output length to 32, the batch size to 32, and the learning rate to 5e-5. We used a beam size of 10 during the training process. We performed a 3:1:1 split for training, validation, and test to evaluate our model. We implemented all training with 1 NVIDIA GeForce RTX H800 GPU with 80GB memory and CUDA $12.2$ on PyTorch. It took about 4 hours for smart contract vulnerabilities detection training. For the inference of large language models, we set the max input length to 2048 and the max output length to 512. To ensure the deterministic output, we set the temperature to 0 and top-p to 1. A repetition penalty of 1.2 was applied to avoid repeated generation. The inference was performed on a server equipped with 2 NVIDIA GeForce RTX H800 GPUs, each with 80GB memory.

\subsection{Experimental Results}
In this section, we present experimental results to answer the research question.

1) RQ1: To evaluate the effectiveness of our proposed SAEL method, we compared it with state-of-the-art baseline methods. The experimental results are presented in Table \ref{comparison}.

For reentrancy vulnerability detection, SAEL achieved the best performance in terms of Precision, Recall, and F1-score, reaching 93.62\%, 94.96\%, and 94.29\%, respectively, significantly outperforming all other baseline methods. ReVulDL, a pre-trained model-based method, was the second-best performer with an F1-score of 92.14\%. For timestamp dependency vulnerability detection, SAEL also demonstrated superior performance, ranking first in all three evaluation metrics with a Precision of 90.16\%, Recall of 95.17\%, and F1-score of 92.60\%, surpassing all baseline methods. Another pre-trained model-based method PSCVFinder achieved the F1-score of 89.76\%, ranking second in this task. This demonstrates the effectiveness of explanations for smart contract vulnerability detection tasks. In the detection of integer overflow/underflow vulnerabilities, SAEL continued to show excellent performance, achieving the highest F1-score of 83.16\%. This surpassed the second-best method, Peculiar, which had an F1-score of 75.14\%. For delegatecall vulnerability detection, SAEL maintained its leading position with an F1-score of 80.31\%, outperforming the next best method, PSCVFinder, which achieved an F1-score of 70.87\%.

\begin{tcolorbox}
\textbf{[RQ1]}: SAEL consistently outperformed 12 state-of-the-art baseline methods across all four types of vulnerabilities (reentrancy, timestamp dependency, integer overflow/underflow, and delegatecall).
\end{tcolorbox}

2) : We explore the influence of various components and features on the performance of smart contract vulnerability detection in SAEL. The framework utilizes three key features: raw smart contract code features (R), explanations generated by LLMs (E), and predictions provided by LLMs (P).

\begin{figure}[!h]
\centerline{\includegraphics[width=0.5\textwidth,height=0.2\textwidth]{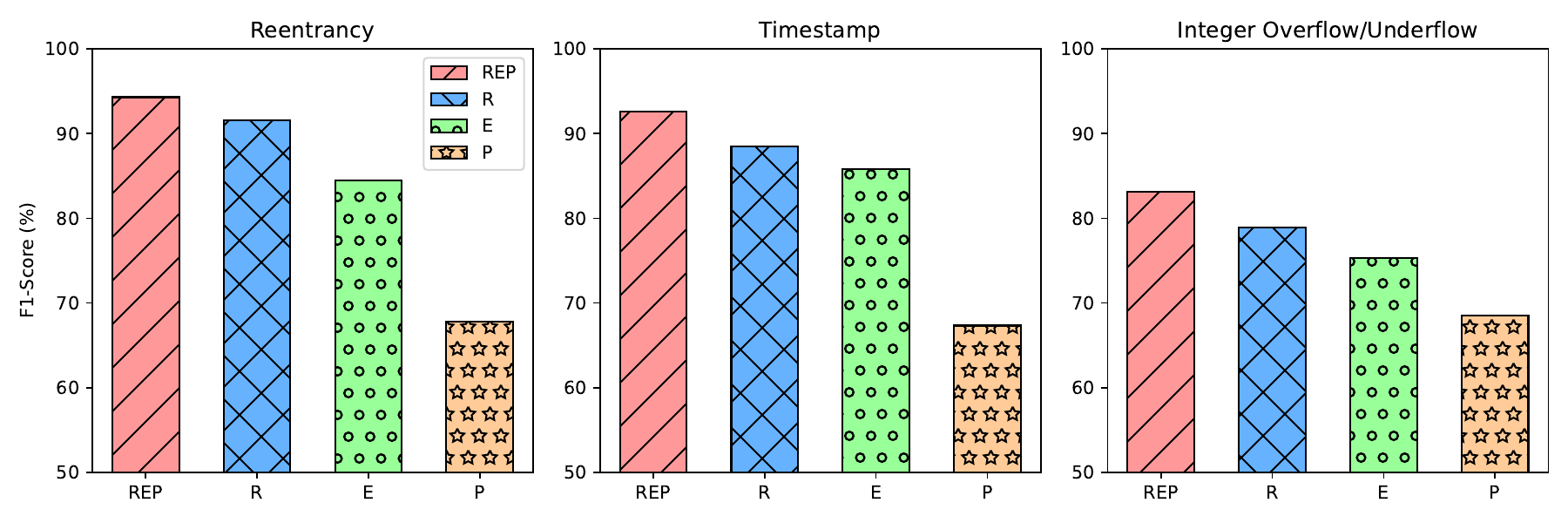}}
\caption{Analysis of Different Features}
\label{feature}
\end{figure}

Fig. \ref{feature} demonstrates the effectiveness of the REP feature, which is obtained by integrating R, E, and P features through Adaptive Mixture-of-Experts. The REP feature achieves the highest F1-scores for all three vulnerability types: 94.29\% for reentrancy, 92.60\% for timestamp dependency, and 83.16\% for integer overflow/underflow. When R, E, and P features are used individually, they result in lower F1-scores across all vulnerability types. The raw code features (R) consistently contribute the most, followed by explanations (E), with predictions (P) having the least impact. This finding suggests that the incorporation of explanations (E) and prediction results (P) generated by LLMs significantly boosts performance compared to using raw code features (R) alone.

The explanations and prediction results generated by LLMs plays a crucial role in the smart contract vulnerability detection task. Detecting vulnerabilities in smart contracts requires a deep understanding of the semantics and context of the code. LLMs, pre-trained on vast amounts of code and natural language data, possess the ability to comprehend code semantics and context. The explanations generated by LLMs highlight potential vulnerabilities and provide the reasons behind the predictions. These explanations offer a high-level understanding of the behavior of the code and potential security risks, compensating for the limitation of lacking semantic understanding when relying solely on raw code features.

\begin{figure}[!h]
\centerline{\includegraphics[width=0.5\textwidth,height=0.2\textwidth]{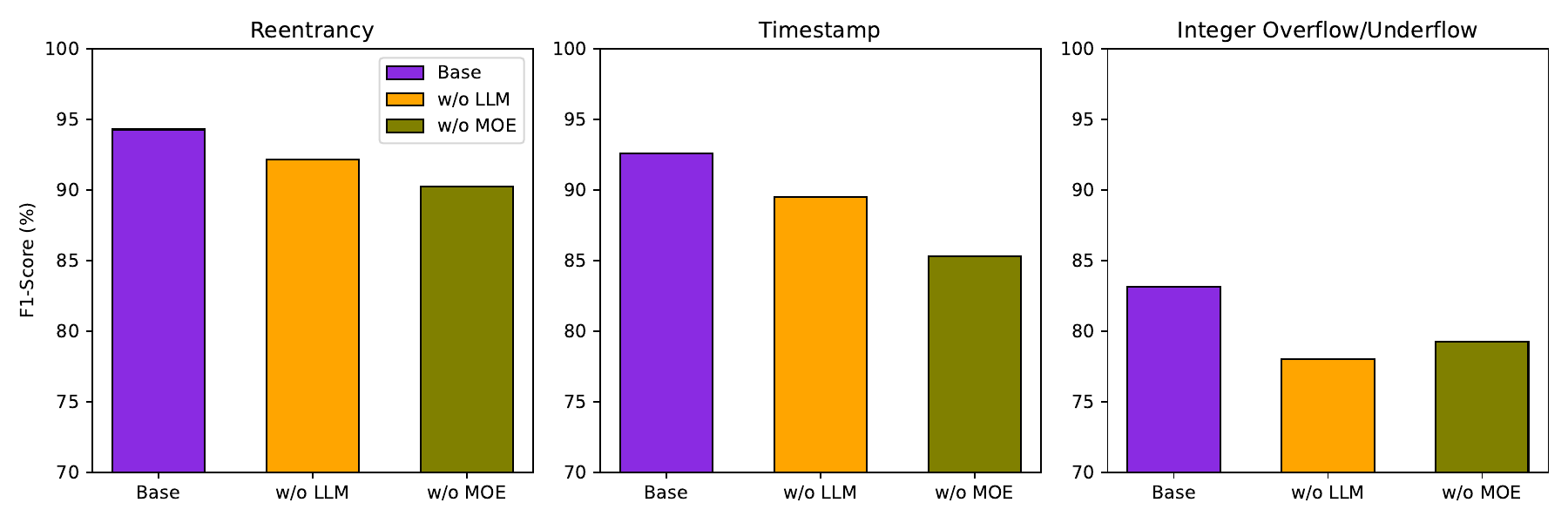}}
\caption{Analysis of Different Modules}
\label{module}
\end{figure}

The impact of different modules on the performance of SAEL is analyzed in Fig. \ref{module}. The complete SAEL model (Base) achieves the best results across all three vulnerability types, with F1-scores of 94.29\% for reentrancy, 92.60\% for timestamp dependency, and 83.16\% for integer overflow/underflow. When the language model module (w/o LLM) is removed, the F1-scores decrease across all vulnerability types. Here, w/o LLM refers to performing prompt-tuning only on the raw code without utilizing LLM-generated explanations and predictions. Further removing the Adaptive Mixture-of-Experts module (w/o MOE) results in a more significant performance drop. The w/o MOE condition indicates direct averaging of the prediction results from the three feature types without dynamic weight adjustment.

These findings highlight the positive contributions of both modules to the overall performance across all vulnerability types. The Adaptive Mixture-of-Experts module appears to have a more pronounced impact, especially for reentrancy and timestamp dependency vulnerabilities, where its removal leads to a larger performance drop compared to removing the LLM module.

\begin{tcolorbox}
\textbf{[RQ2]}: The explanations and predictions generated by LLMs significantly enhance vulnerability detection performance. The Adaptive Mixture-of-Experts module further optimizes detection by dynamically adjusting feature weights.
\end{tcolorbox}

\begin{figure}[!h]
\centerline{\includegraphics[width=0.5\textwidth,height=0.2\textwidth]{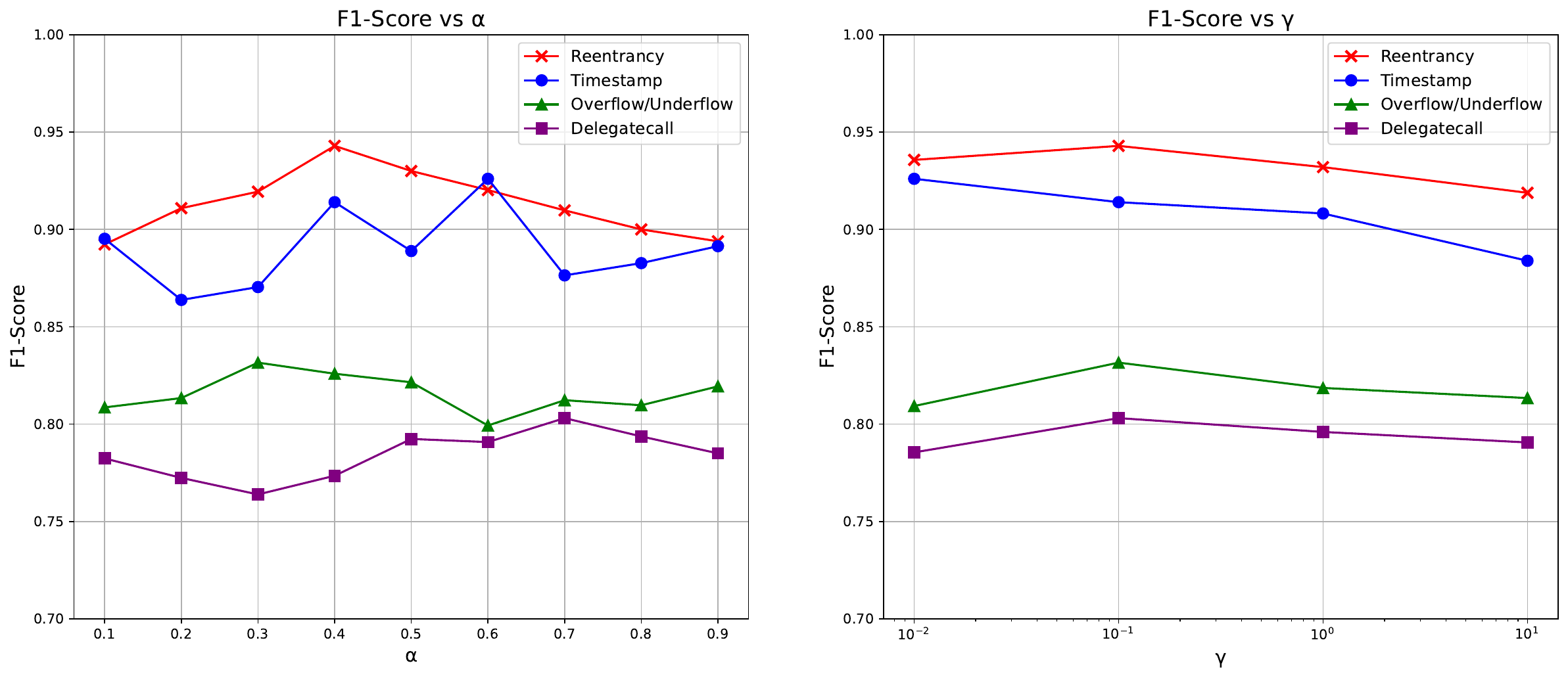}}
\caption{Performance of SAEL with different parameters.}
\label{sensitive}
\end{figure}

3) RQ3: Based on the experimental results shown in Fig. \ref{sensitive}, our SAEL exhibits parameter sensitivity across four types of vulnerabilities: reentrancy, timestamp dependency, integer overflow/underflow, and delegatecall. As the two key parameters $\alpha$ and $\gamma$ in Adaptive Mixture-of-Experts vary, the F1-scores for these vulnerability types show different trends.

The parameter $\alpha$ balances the feature weight adjustment loss $L_{feature}$ and the overall cross-entropy loss $L_{pred}$ in the loss function design. For reentrancy and timestamp dependency vulnerabilities, the F1-scores peak when $\alpha$ is around 0.4-0.6, indicating that a moderate balance between the two losses improves detection performance. Integer overflow/underflow vulnerabilities show a relatively stable performance across different $\alpha$ values, while delegatecall vulnerabilities exhibit more fluctuations.

Furthermore, the parameter $\gamma$ controls the proportion of weight regularization loss Lreg in the total loss to prevent rapid changes in feature weights. When $\gamma$ is between $10^{-2}$ and $10^{0}$, the detection performance for all vulnerability types is relatively stable, indicating that moderate weight regularization aids convergence and generalization. However, as $\gamma$ increases beyond $10^{0}$, the model performance generally declines across all vulnerability types, with reentrancy and timestamp dependency showing the most significant drops. This suggests that these two vulnerability types require more flexible feature weight adjustments, which are hindered by excessive regularization. Integer overflow/underflow and delegatecall vulnerabilities show less sensitivity to high $\gamma$ values, maintaining relatively stable performance even at $\gamma$ = $10^{1}$. This could indicate that the detection of these vulnerabilities relies more on consistent, general patterns that are less affected by strict weight regularization.

\begin{tcolorbox}
\textbf{[RQ3]}: Regarding to the four vulnerabilities, the loss function balance parameter $\alpha$ and weight regularization parameter $\gamma$ in Adaptive Mixture-of-Experts have a significant impact on model performance and need to be properly adjusted.
\end{tcolorbox}

4) RQ4: To evaluate the zero-shot performance of SAEL on smart contract vulnerability detection, we randomly selected 400 labeled samples from various datasets: 100 contracts for reentrancy vulnerability detection from the Ethereum Smart Contract Dataset \cite{zhuang2020smart}, and 300 contracts from the Ethernet Open Dataset \cite{liu2023rethinking} (100 each for timestamp dependency, integer overflow/underflow, and delegatecall vulnerability detection). All selected contracts were excluded from the train and test datasets. The ratio of vulnerable to non-vulnerable contracts for each vulnerability type is as follows: reentrancy (32:68), timestamp dependency (43:57), and both integer overflow/underflow and delegatecall (22:78).

\begin{figure}[!h]
\centerline{\includegraphics[width=0.5\textwidth,height=0.25\textwidth]{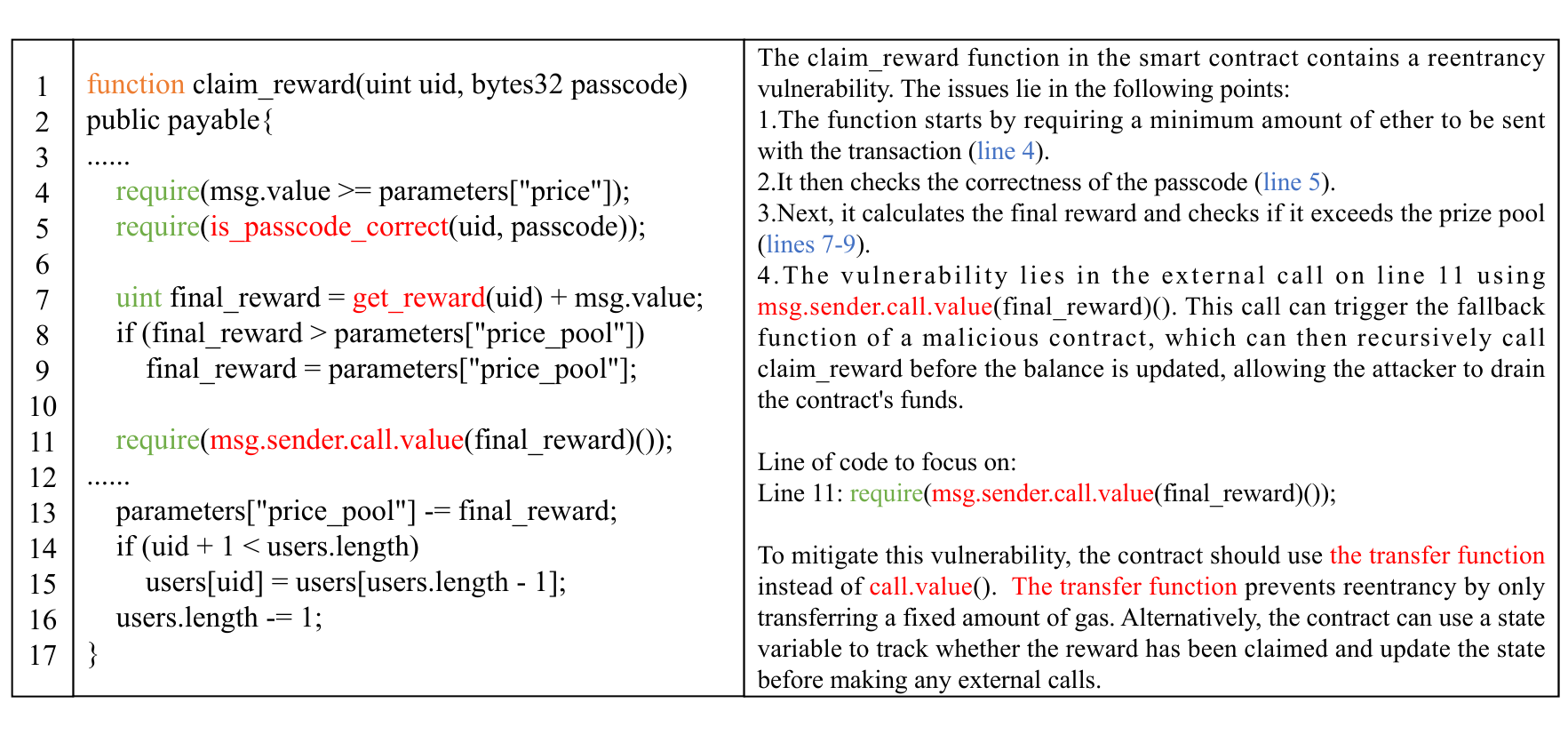}}
\caption{A simplified real-world smart contract with a complex reentrancy vulnerability and its explanation generated by SAEL.}
\label{case}
\end{figure}

We show a simplified case in Fig. \ref{case}. The reentrancy vulnerability in this contract differs from traditional ones in the following aspects: the triggering conditions are more hidden, requiring password checks and prize pool condition judgments; exploiting the vulnerability requires cross-contract interactions; complex reward calculation logic may interfere with the ability of detection tools to identify key vulnerability points; the location of contract state updates is uncommon, increasing detection difficulty. Usually reentrancy vulnerabilities are caused by updating the balance status after the transfer, but in this case the bonus pool balance is deducted before deleting the user information. Traditional tools based on rule matching and program analysis including Oyente \cite{luu2016making}, Securify \cite{tsankov2018securify}, Smartcheck \cite{tikhomirov2018smartcheck}, Slither \cite{feist2019slither} and Mythril \cite{mueller2017mythril} fail to accurately detect such complex vulnerabilities, mainly due to the difficulty of rule matching in covering all variants, the difficulty of program analysis in understanding code semantics, and complex logic interfering with vulnerability localization. As shown in Table \ref{zero_shot}, SAEL demonstrated strong vulnerability detection capabilities in a zero-shot manner.

\begin{table}
\centering
\caption{The results of SAEL in a zero-shot manner.}
\label{zero_shot}
\begin{tabular}{cccc}
\hline
Vulnerabilities & Precision(\%) & Recall(\%) & F1(\%) \\
\hline
Reentrancy & 93.50 & 90.60 & 92.00 \\
Timestamp & 88.60 & 90.70 & 89.60 \\
Overflow/Underflow & 85.70 & 81.80 & 83.70 \\
Delegatecall & 85.00 & 77.30 & 81.00 \\
\hline
\end{tabular}
\end{table}

SAEL addresses the above challenges by leveraging LLMs for in-depth code understanding and combining them with the Adaptive Mixture-of-Experts to dynamically adjust different features. Specifically, LLMs can deeply understand code semantics, accurately grasping vulnerability triggering conditions. Adaptive Mixture-of-Experts enables flexible capture of key features which can adapt to complex vulnerability scenarios. Furthermore, the analysis generated by SAEL provide detailed explanations of vulnerability principles.

\begin{tcolorbox}
\textbf{[RQ4]}: Our findings demonstrate the strong zero-shot capability of SAEL. Through a case study, we showcase the superior ability of SAEL to detect complex vulnerabilities compared to rule-based methods, while also providing the comprehensive explanation.
\end{tcolorbox}

\section{Related Work}

% In this section, we review three key areas of related work for smart contract vulnerability detection: rule-based methods, pre-trained model approaches, and LLM-based methods.

In this section, we review four key areas of related work for smart contract vulnerability detection and related applications: rule-based methods, pre-trained model approaches, LLM-based methods, and applications of LLMs in broader software engineering tasks.

\subsection{Rule-based Methods}

Various studies have employed traditional program analysis methods to identify particular vulnerabilities in smart contracts. Oyente \cite{luu2016making} leverages symbolic execution to uncover four types of vulnerabilities by exploring different execution paths within smart contracts. Mythril \cite{mueller2017mythril} and SmartCheck \cite{tikhomirov2018smartcheck} rely on pattern matching techniques to detect vulnerabilities based on a set of predefined rules. Securify \cite{tsankov2018securify} utilizes formal verification using logical languages to ensure the security of smart contracts. Osiris \cite{torres2018osiris} and Manticore \cite{mossberg2019manticore} combine symbolic execution and taint analysis, while Maian \cite{tann2018towards} uses symbolic analysis and concrete validation. Slither \cite{feist2019slither} is a static analysis framework that employs data flow and taint analysis. 

\subsection{Pre-trained models-based Methods}

Several studies have leveraged pre-trained models in various ways to enhance smart contract security. Peculiar \cite{wu2021peculiar} focused on improving generalization through pre-training, and ReVulDL \cite{zhang2022reentrancy} utilized a graph-based pre-training model to capture propagation chain relationships. PSCVFinder \cite{yu2023pscvfinder} utilizes prompt-tuning to bridge the gap between pre-training task and smart contract vulnerability detection task.

\subsection{LLM-based Methods}

Recent studies \cite{chen2023chatgpt, david2023you} have evaluated the performance of LLMs on real-world datasets, revealing that LLMs encounter performance-related challenges due to a high prevalence of false positives. Hu et al. \cite{hu2023large} investigated the application prospects of LLMs in smart contract vulnerability detection from new perspectives. Sun et al. \cite{sun2024gptscan} introduced GPTScan, the first tool that combines GPT with program analysis for detecting logic vulnerabilities in smart contracts. GPTScan breaks down each logic vulnerability type into scenarios and properties, utilizes GPT to match candidate vulnerabilities, and then confirms them through static analysis. Ma et al. \cite{ma2024combining} introduced iAudit, a two-stage framework leveraging large language models for detecting vulnerabilities and providing explanations. These works demonstrate the great potential of LLMs in the field of smart contract security. 

\subsection{Applications of LLMs in Software Engineering}

Large Language Models (LLMs) have been increasingly applied to various software engineering tasks. For code review automation, Lu et al.\cite{lu2025deepcrceval} introduced DeepCRCEval, integrating LLMs to improve evaluation quality and efficiency. Lu et al.\cite{lu2023llama} proposed LLama-Reviewer, which uses parameter-efficient fine-tuning for effective and resource-efficient code review. In code understanding, Shen et al.\cite{shen2024dependency} designed a dependency-aware framework for method naming and consistency checking, showing LLMs can enhance these tasks through advanced sampling strategies. For issue resolution, Zan et al.\cite{zan2024swe} developed SWE-bench-java, a benchmark for evaluating LLMs on GitHub issue resolution, demonstrating LLMs’ potential in automated software maintenance. These works reflect the versatility and effectiveness of LLMs in modern software engineering.

% However, they require require a costly payment to use the APIs of close-sourced LLMs and have not fully utilized explanations to enhance the performance of vulnerability detection.

% In contrast, our SAEL is the first work to leverage LLMs to offer explanations for smart contract vulnerabilities, enhancing the interpretability of the smart contract vulnerability detection.

\section{Threats to Validity}

\textbf{Internal Validity}: Recent research indicates that the precise influence of hyperparameters on the performance of LLMs and Deep Learning models remains unclear \cite{raffel2020exploring, akiba2019optuna, guo2024exploring}. In our work, we have applied the Tree-structured Parzen Estimator (TPE) \cite{bergstra2011algorithms} to enhance our model performance. However, we recognize that alternative configurations could potentially deliver comparable or better results. Consequently, we plan to explore further settings in our subsequent research.

\textbf{External Validity}: The SAEL model requires substantial labeled training data to extract sufficient features, which may restrict its ability to detect new categories of vulnerabilities where traditional approaches may prove more effective. To overcome this limitation, we can reduce the reliance on labeled data by continuing the pretraining task on unsupervised datasets of smart contracts.

\section{Conclusion}

In this paper, we propose SAEL, a smart contract vulnerability detection approach based on LLMs. Unlike the prior work, SAEL novelly utilizes explanations generated by general-purposed LLMs as a feature to enhance the performance of smart contract vulnerability detection. Furthermore, we introduce Adaptive Mixture-of-Experts to dynamically adjust the weights of prediction results for LLMs, explanation features, and contract code features. Our approach outperforms state-of-the-art methods. For future work, we aim to decrease our dependency on labeled data by furthering the pretraining process using unsupervised datasets of smart contracts.

\section{Acknowledgement}
This work was supported by the Alliance of International Science Organizations Collaborative Research Program (No.ANSO-CR-KP-2022-03).

\bibliographystyle{IEEEtran}
\balance
\bibliography{IEEEabrv,References}

\end{document}